\documentstyle[epsfig,preprint,aps,prb]{revtex}
\draft

\begin{document}

\title{Effects of biquadratic exchange on the spectrum of elementary 
excitations in spin ladders}

\author{S. Brehmer$^1$, H.-J. Mikeska$^1$, M. M\"uller$^1$, N. Nagaosa$^2$, S. Uchida$^2$} 
\address{$^1$Institut~f\"ur~Theoretische~Physik, Universit\"at~Hannover, 
30167~Hannover, Germany\\
$^2$Department of Applied Physics, University of Tokyo, Bunkyo-ku,
Tokyo 113, Japan}

\maketitle

\begin{abstract}
We investigate the influence of biquadratic exchange interactions
on the low-lying excitations of a $S=\frac{1}{2}-$ladder using perturbation
theory, numerical diagonalization of finite systems and exact results 
for ladders with matrix product ground states. We consider in particular 
the combination of biquadratic exchange interactions corresponding to
ring exchange on the basic ladder plaquette. We find that a moderate amount
of ring exchange reduces the spin gap substantially and makes equal
bilinear exchange on legs and rungs consistent with experimentally
observed spectra.  
\end{abstract}

\medskip

\pacs{PACS numbers: 75.10Jm}

\section{Introduction}

Two-legged spin ladders with spins 1/2 have attracted considerable
interest over the past years, both as ideal models for quasi 1D
materials and as a theoretical model for a spin liquid characterized
by an excitation gap. From the geometric structure of the two-legged
spin ladder as shown in fig. 1(a) it is clear that the main exchange
interactions are expected along the legs and across the rungs; if the
most general interaction for a plaquette formed by four spins
$S=\frac{1}{2}$ on two neighboring rungs is considered, exchange along
the two diagonals (corresponding to next nearest neighbor
interactions) and biquadratic exchange interactions appear in addition.

This general plaquette hamiltonian for S=1/2 is formulated as:

\begin{eqnarray}
H &=&  \sum_n \{ J_{rung} \: {\mathbf S}_{1,n} {\cdot\mathbf S}_{2,n}     
     + J_{leg} \: ({\mathbf S}_{1,n} {\cdot\mathbf S}_{1,n+1}
                 + {\mathbf S}_{2,n} {\cdot\mathbf S}_{2,n+1}) 
     \nonumber \\
  &+& J_{diag} \: ({\mathbf S}_{1,n} {\cdot\mathbf S}_{2,n+1}
                 + {\mathbf S}_{2,n} {\cdot\mathbf S}_{1,n+1}) 
     \nonumber \\
  &+& V_{RR} \: ({\mathbf S}_{1,n} {\cdot\mathbf S}_{2,n}) 
                ({\mathbf S}_{1,n+1} {\cdot\mathbf S}_{2,n+1})  
    + V_{LL} \: ({\mathbf S}_{1,n} {\cdot\mathbf S}_{1,n+1}) 
                ({\mathbf S}_{2,n} {\cdot\mathbf S}_{2,n+1})    
     \nonumber \\  
  &+& V_{DD} \: ({\mathbf S}_{1,n} {\cdot\mathbf S}_{2,n+1}) 
                ({\mathbf S}_{2,n} {\cdot\mathbf S}_{1,n+1})\}.  
\end{eqnarray}

For simplicity we have assumed equal exchange interactions $J_{leg}$
on the two legs and $J_{diag}$ on the two diagonals, a natural
assumption for the symmetric structure of fig. 1 which we will
consider in the following. Recently theoretical studies have
demonstrated that this generalized hamiltonian (i) allows to formulate
models interpolating smoothly between the dimer and Haldane limits for
the ground state \cite{KolM97} and (ii) has a parameter space
sufficiently large to allow to study quantum phase transitions
\cite{KolM98}. The typical hamiltonian for both applications has some
nonvanishing biquadratic terms. In the past, only little attention
has been paid to biquadratic exchange terms, although finite strength
of these terms was e.g.\ found in the spectra of small clusters of
magnetic ions
\cite{FalFKG87}.

Recently, the possible importance of biquadratic exchange for some
properties of low-dimensional spin systems has been pointed out: Honda
et al.\ \cite{HonKW93} have argued for finite ring exchange
(corresponding to a particular combination of biquadratic terms) on
the basic $\rm CuO_2$ plaquette. The relevance of ring exchange for
magnetization plateaus has been discussed for the spin ladder by
Sakai and Hasegawa \cite{SakH98} and for solid $\rm He^3-$films by
Momoi et al.\ \cite{MomSK98}. In this paper we study the effect of
the biquadratic terms in the hamiltonian (1) on the low-lying
excitations of this 'generalized spin ladder' by both analytical and
numerical approaches. The motivation for this work is twofold:

Firstly, we want to extend the knowledge obtained analytically in
ref. \onlinecite{KolM98} to a wider range of strengths of the
biquadratic interactions.

Secondly, we want to investigate the relevance of ring exchange of
finite strength $J_{ring}$ as introduced by Honda et
al. \cite{HonKW93} on the determination of coupling constants from
ladder spectra. The standard analysis of experimental data on quasi 1D
ladder systems starts from eq. (1) with $J_{diag}$ and all biquadratic
couplings set equal to 0 and results in $J_{leg} \approx 2 \:
J_{rung}$ from inelastic neutron scattering experiments on $\rm
Sr_{14}Cu_{24}O_{41}$ \cite{EccUAEMU98} and $\rm
La_6Ca_8Cu_{24}O_{41}$ \cite{MatE98}, as well as from NMR experiments
on these substances \cite{TakMEU98,MagMKIA98,ImaTSHC98}. This large
value for the ratio $J_{leg} / J_{rung}$ is not understood at present:
It is not expected from the geometric structure of the ladder, from
electronic structure calculations \cite{MulARDS98} a value somewhat
larger than unity is obtained and inter-ladder interactions
\cite{MiyTJU98} which have recently been considered, do not
resolve the discrepancy.

In the following we show that even a moderate amount of ring exchange
$J_{ring}$ is relevant for the determination of the coupling constants
in the two leg ladder from the energies of elementary excitations. We
define the strength $J_{ring}$ on a plaquette of four spins $i,j,k,l$
in terms of the permutation operator $P_{ijkl}$ by the symmetric
contribution

\[H_{ring} \:=\: \frac{1}{2} J_{ring} 
      \sum_{plaquettes} \left( P_{ijkl} + P_{lkji} \right) \] 
to the hamiltonian. Since we deal with spins
$S=1/2$, $H_{ring}$ is identical to a special choice of constants in
eq. (1), namely

\[V^{ring}_{RR} = V^{ring}_{LL} = - V^{ring}_{DD} = 2 J_{ring}, \: \: 
      J^{ring}_{rung} = J_{ring}, \:\: J^{ring}_{leg} = J^{ring}_{diag} 
      = \frac{1}{2} J_{ring}. \]
A natural hamiltonian for the geometric structure of the spin ladder
then is obtained by bilinear exchange terms and ring exchange terms, 
leading to the hamiltonian of eq. (1) with 

\begin{eqnarray}
J_{rung} &=& J_{rung}^{bl} + J_{ring}, \: 
J_{leg}  = J_{leg}^{bl} + \frac{1}{2} J_{ring}, \: 
J_{diag} =  \frac{1}{2} J_{ring}   \nonumber \\ 
V_{RR} &=& V_{LL} = - V_{DD} = 2 J_{ring}.
\end{eqnarray}

In the following we will present results on the influence of $J_{ring}
\ne 0$ on the dispersion of low-lying excitations for
antiferromagnetic ladders using perturbation theory (section II),
numerical calculations covering the case of experimental interest
(section III) and discussing the relevance for the dispersion of
exactly known excitation branches (in the case of special choices for
the strength of biquadratic exchange).

\medskip

\section{Perturbation theory}

We have calculated the dispersion of the lowest triplet excitation in an
expansion in the neighbourhood of the dimer point (where $J_{rung} =1$ is
the only nonvanishing exchange constant) \cite{Mul98}. In the following we
restrict the discussion to the hamiltonian of experimental interest

\begin{eqnarray}
H \:&=&\: \sum_n \{ {\mathbf S}_{1,n} {\cdot\mathbf S}_{2,n}     
     + J_{leg} \: ({\mathbf S}_{1,n} {\cdot\mathbf S}_{1,n+1}
                 + {\mathbf S}_{2,n} {\cdot\mathbf S}_{2,n+1})\}
     \:+\: H_{ring}.
\end{eqnarray}       
    
The result for the spectrum of the lowest triplet excitation 
to third order in $J_{leg}, J_{ring}$ is

\begin{eqnarray}       
\omega_{S=1}(k) \:&=&\: 
  \left( t_0 \:+\: t_1 \: \cos\:k \:+\: t_2 \: \cos\:2k 
            \:+\: t_3 \: \cos\:3k \right)  \\
t_0 \:&=&\: 1 \:-\: 2 J_{ring} 
              \:+\: \frac{3}{4} \left(J_{leg} - J_{ring}\right)^2
              \:+\: \frac{3}{8} \left(J_{leg} - J_{ring}\right)^2 
                       \left(J_{leg} + 5 J_{ring}\right) \nonumber   \\
t_1 \:&=&\: J_{leg} \:+\: J_{ring} \:-\:
          \frac{1}{4} \left(J_{leg} - J_{ring}\right)^2 
                      \left(J_{leg} + J_{ring}\right) \nonumber \\ 
t_2 \:&=&\: -\frac{1}{4} \left(J_{leg} - J_{ring} \right)^2
            -\frac{1}{4} \left(J_{leg} - J_{ring} \right)^2
                \left(J_{leg} + J_{ring}\right)  \nonumber \\ 
t_3 \:&=&\: \frac{1}{8} \left(J_{leg} - J_{ring} \right)^2
                \left(J_{leg} + J_{ring}\right)  \nonumber  
\end{eqnarray}       

The leading coefficient for the dispersion is 
$t_1 \propto J_{leg} \:+\: J_{ring}$ in first order. This perturbative
result shows: When $J_{ring} > 0$ is present, an analysis
of the spectra in terms of bilinear exchange only will
lead to an effective value of $J_{leg}$ which is increased in
comparison to the value found for $J_{ring} = 0$.
We will see in the next section from numerical diagonalization that
this result of perturbation theory continues to be qualitatively true
for larger values of $J_{leg}$.

\section{The symmetric ladder with biquadratic exchange}

Motivated by the results of recent inelastic neutron scattering
experiments on the quasi 1D ladder material $\rm
La_6Ca_8Cu_{24}Sr_{41}$ \cite{MatE98} we have calculated numerically
excitation spectra for spin ladders with the hamiltonian of eq. (2)
using periodic boundary conditions for a finite number of spins. We
have used the Lanczos method for ladders with a total of 24 spins
i.e.\ 12 rungs. Because of space inversion the number of wavevectors
$k$ with different energies is 7, $k=p \pi/6, \: p = 0,1 \ldots 6$. In
figs. 2a-d we present the results for the lowest negative parity
excitation (parity with respect to interchange of the legs).  This is
the lowest excitation with the ladder gap at $k=\pi$; it is very
likely a triplet excitation, since it is found with identical energies
for both $S^z_{tot}=1$ and $S^z_{tot}=0$ ($S_{tot} > 1$, however, is
not formally excluded). We show the variation of the dispersion with
$J_{ring}$ between the limits $-0.3 \dots +0.3$ for four values of
$J_{leg}$. It should be noted that the unit of energy is the bilinear
rung exchange $J_{rung}^{bl}$ (without the effective contribution
$J_{rung}^{ring}$ resulting from the ring exchange). $J_{rung}^{bl}$
should be fixed by the magnitude of the gap.

For a simple quantitative presentation of the effect of ring exchange
on the spectra we have considered the ratio 

\begin{equation}
R = \frac{\omega({2\pi \over 3})}{\omega({\pi)}}. 
\end{equation}

This quantity may conveniently serve for comparison with experimental
data since the most accurate spectra are obtained for $k >
\frac{\pi}{2}$ and it is here where $\omega(k)$ varies most.  The
curves in fig. 3 show the dependence of the ratio $R$ on $J_{ring}$ for
$J_{leg} / J_{rung} = 0.5, 1.0, 1.5, 2.0$.  The analysis of the
experiments gives $R \approx 5$, which may be realized for different
combinations of $J_{leg}$ and $J_{ring}$: Two examples are $J_{leg}
\approx 1.8, \: J_{ring} \approx 0$ and $J_{leg} \approx 1.0, \:
J_{ring} \approx 0.14$. Thus we conclude from the data shown in fig. 2
the tendency that a
finite $J_{ring}>0$ tends to simulate the effect of an increased
bilinear exchange $J_{leg}$ on the legs. This is 
in agreement with  perturbation theory as discussed in sect. II. We 
have checked that the perturbative result of eq. (3) is a good 
approximation to the numerical data for $J_{leg} = 0.5$ for 
$J_{ring} < 0.1$; for larger ring exchange a transition to a new
phase occurs (see below) such that perturbation theory does no more
apply.
 
The main physics behind the importance of $J_{ring}$ for the
determination of bilinear exchange constants is the following: the
influence of $J_{ring}$ on the gap (at $k=\pi/2$) is relatively much
stronger than on the dispersion at intermediate wavevectors. Actually,
for $J_{leg} < 1$, the gap appears to vanish for a critical magnitude
of $J_{ring}$, indicating a phase transition. This phase transition
may also be responsible for the irregular behaviour of the dispersion
curves for $J_{leg} = 0.5, J_{ring} = 0.3$; this system may actually
be in a different phase. It is the neighbourhood of this phase
transition also for values $J_{leg} \propto O(1)$ which explains the
strong influence of the biquadratic exchange terms. We expect this
phase transition to be of the same type as discussed before for
generalized spin ladders \cite{NerT97,KolM98a,KolM98}, leading to a
spontaneously dimerized ground state.

The energy of the spin gap is the basic input for the determination of
the energy scale in ladder systems. Owing to the large effect of
finite $J_{ring}$ on the spin gap the values of the exchange constants
in $\rm CuO_2-$planes will have to be renormalized considerably. For
$J_{ring} \approx 0.14 \: J_{rung}, \: J_{leg} \approx J_{rung}$ the
spin gap is $\Delta \approx 0.28 \: J_{rung}$ (compared to $\Delta
\approx 0.51 \: J_{rung}$ in the case $J_{ring} = 0$). This implies an
energy scale which is larger by a factor close to 2 owing to the ring
exchange terms. Thus the basic exchange energy $J_{rung}$ between two
Cu ions changes from $\approx 800 K$ to $\approx 1400 K$, comparable
to the magnitude of the main exchange constant in the 2D material $\rm
La_2CuO_4$.

If the $k-$dependent dispersion is considered in more detail, the
situation is of course more complex: In particular the presence or
absence of a maximum between $k=0$ and $k = \pi$ is affected by the
value of $J_{ring}$. At present, however, this is a point of little
relevance for experimental results. In fig. 2b we have also included
results for the lowest excitation energy with positive parity for
$J_{ring} = 0, \pm 0.3$: these data illustrate our general observation
that the effect of biquadratic exchange on the positive parity
excitation energies is less spectacular.

\section{Exact dispersion relations for spin ladders with special values of 
the exchange constants}

It was observed in ref. \onlinecite{KolM98} that the general plaquette
hamiltonian for the ladder structure allows some combinations of
parameters which result in either a rung dimer (singlet) ground state
with exactly known triplet excitations or a Haldane-liquid like ground
state with exactly known singlet excitations, both with negative
parity. In both cases the dispersion is a pure cosine dispersion. The
physical picture of these exact excited states is simple: An excited
rung triplet (singlet) propagates to the neighboring rungs both on its
right and left side without exciting states containing two or three
quanta. The interesting and natural question, however, whether these
exact excitations are of interest, in particular whether they are the
lowest excitations in the corresponding ladder, can only be answered
by numerical methods.

The condition for an exact dimer ground state with exact triplet
excitations for the hamiltonian of eq. (1) according to ref. 
\onlinecite{KolM98} reads

\begin{equation}
J_{leg} - J_{diag} = \frac{1}{4} \: (V_{LL} - V_{DD})
\end{equation}

It has to be supplemented \cite{KolM98} by the inequalities which
guarantee the stability of the dimer ground state. For the hamiltonian
of eq. (1,2) the conditions reads

\[ J_{leg}^{bl} \:=\: J_{ring} \:\le\: \frac{1}{5} \]

We illustrate this situation in fig. 4 where we plot the lowest
negative parity excitation in the subspace $S^z_{tot} = 1$ for
$J_{leg}^{bl} = 0.15$ and increasing values of $J_{ring}$. We find that for
$J_{ring} = J_{leg}^{bl} =0.15$ the exact excitation is indeed the lowest
excited state for $S^z_{tot} = 0$ as well as for $ \pm 1$. The
numerical spectra reproduce perfectly the exact dispersion law. Beyond
the stability limit the exact excitation energy is between two highly
excited states (since the rung dimer state then has an energy above
the new ground state proportional to the number of rungs) and is
therefore no more of interest.

It is interesting to note that an exact dimer ground state with an
exact triplet excitation is also realized in 2D and 3D structures: The
condition is that eq.(6) is fulfilled separately for
the exchange interactions in each spatial direction.

A second example where the exact ground state and exact excitation
spectra can be explicitly given is the generalized Bose-Gayen model as
introduced in section V of ref.\onlinecite{KolM98}. This model
is defined by the following choice of constants in the hamiltonian of
eq. (1):

\begin{eqnarray}
J_{rung} &=& y_1, \: \:  J_{leg } = 1, \: \:  J_{diag} = y_2 \nonumber \\
V_{RR} &=& 0, \: \: V_{LL} = \frac{4}{5} (3 - 2y_2), \: \: 
         V_{DD} = \frac{4}{5} (3 y_2 -2).  
\end{eqnarray} 

Varying the two parameters $y_1, y_2$ a number of phases,
distinguished by different ground states are realized, among them the
rung dimer and the valence bond (AKLT) phases. The phase diagram is
reproduced in fig. 5. In ref. \onlinecite{KolM98} exact triplet
(singlet) excitations in the rung dimer (AKLT) phase with negative
parity have been given. We have investigated by exact diagonalization
of ladders with 24 spins to what extent these exact excitations are
the lowest ones for representative points in the phase diagram as
indicated in fig.\ 5. Typical results of these calculations are shown
in fig.\ 6. We find that close to the phase boundary of both the rung
dimer (fig. 6a) and the AKLT (fig.6b) phase the exact excitation is
the lowest one for wavevectors in some neighbourhood of $k = \pi$
(wavevector of the gap). This implies that they are exactly known
critical modes as strongly suggested by physical intuition. Sufficiently 
far away in phase space from the transition lines, however, a
crossover occurs and there may exist lower modes of both negative and
positive parity (fig.6c). On the line $y_2 = 1$ the exact excitations
are dispersionless; this is reproduced for all wavevectors for the
point $(y_1=4, y_2=1)$, whereas for the point $(y_1=-4, y_2=0.95)$ the
deviation from the exact solution is seen to occur for small
wavevectors only.
 
\section{Conclusions}

We have investigated biquadratic exchange interactions in their
influence on the low-lying excitations of $S=\frac{1}{2}-$ladders with
various coupling constants. From both analytical and numerical
calculations we conclude that a quantum phase transition (which is
likely to lead to a spontaneously dimerized ground state) occurs for
small values of the ring exchange $J_{ring} > 0$. At this phase
transition the spin gap vanishes, therefore even small values of
$J_{ring}$ imply large variations of the spin gap and have a strong
influence on the determination of exchange parameters in the ladder in
general. Quantitatively, the presence of a small amount of ring
exchange is shown to be consistent with $J_{rung} \approx J_{leg}$ as
suggested by the geometrical structure of the ladder and with a value
of this basic exchange constant between two neighbouring Cu ions which
is close to the one found in the two-dimensional material $\rm
La_2CuO_4$.

\section*{acknowledgements}

We wish to thank C. Waldtmann for his assistance with the numerical
calculations using the Lanczos algorithm. The work at Hannover U was
supported by the German Ministry for Research and Technology under 
the contract 03MI5HAN5.

\begin{figure}
\centering\epsfig{file=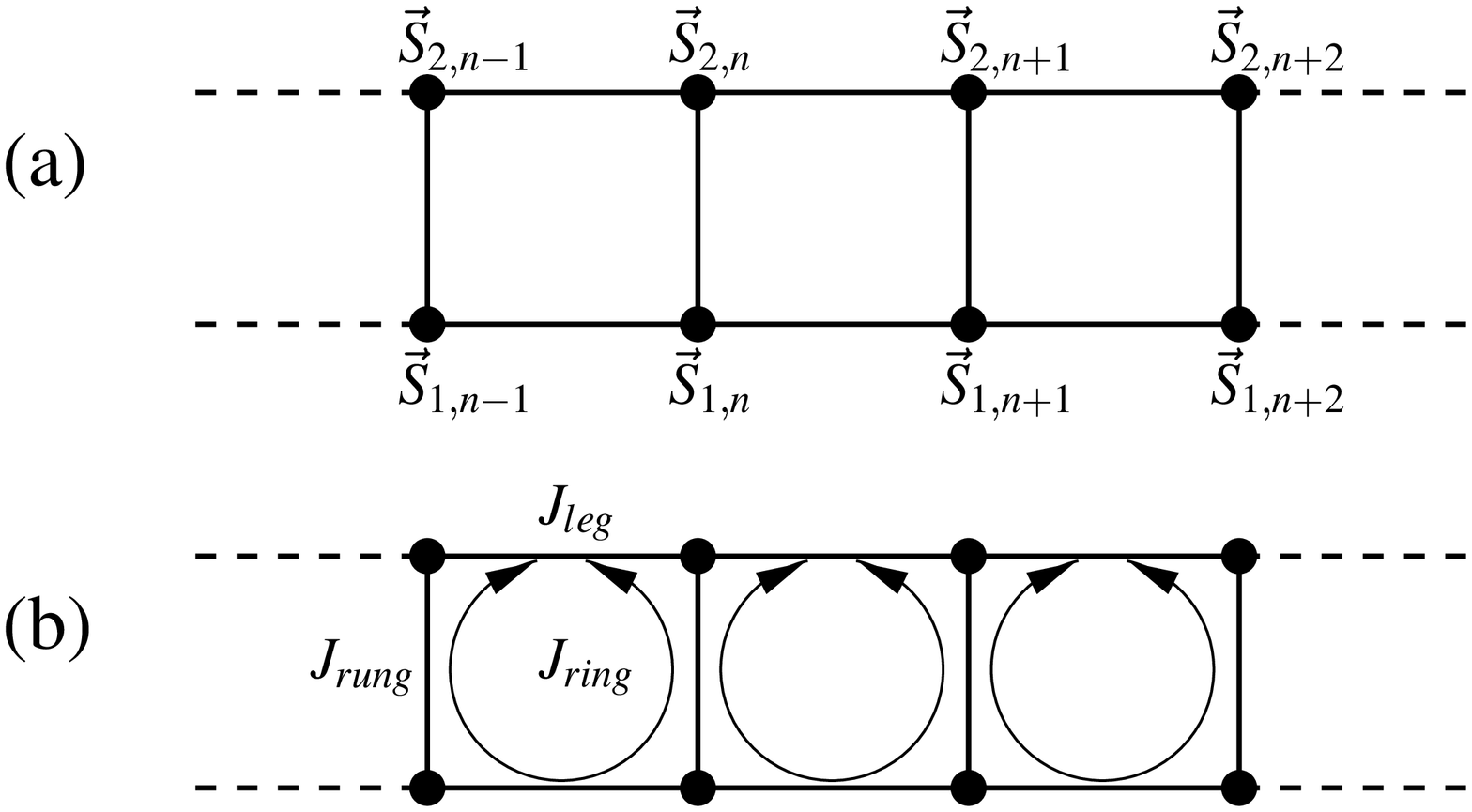,scale=0.57,angle=90}
\vspace{0.3cm}
\caption{(a) Structure of the two leg ladder, (b) possible ring exchange 
around the basic plaquette.} 
\label{ladder}
\end{figure}

\begin{figure}
\centering\epsfig{file=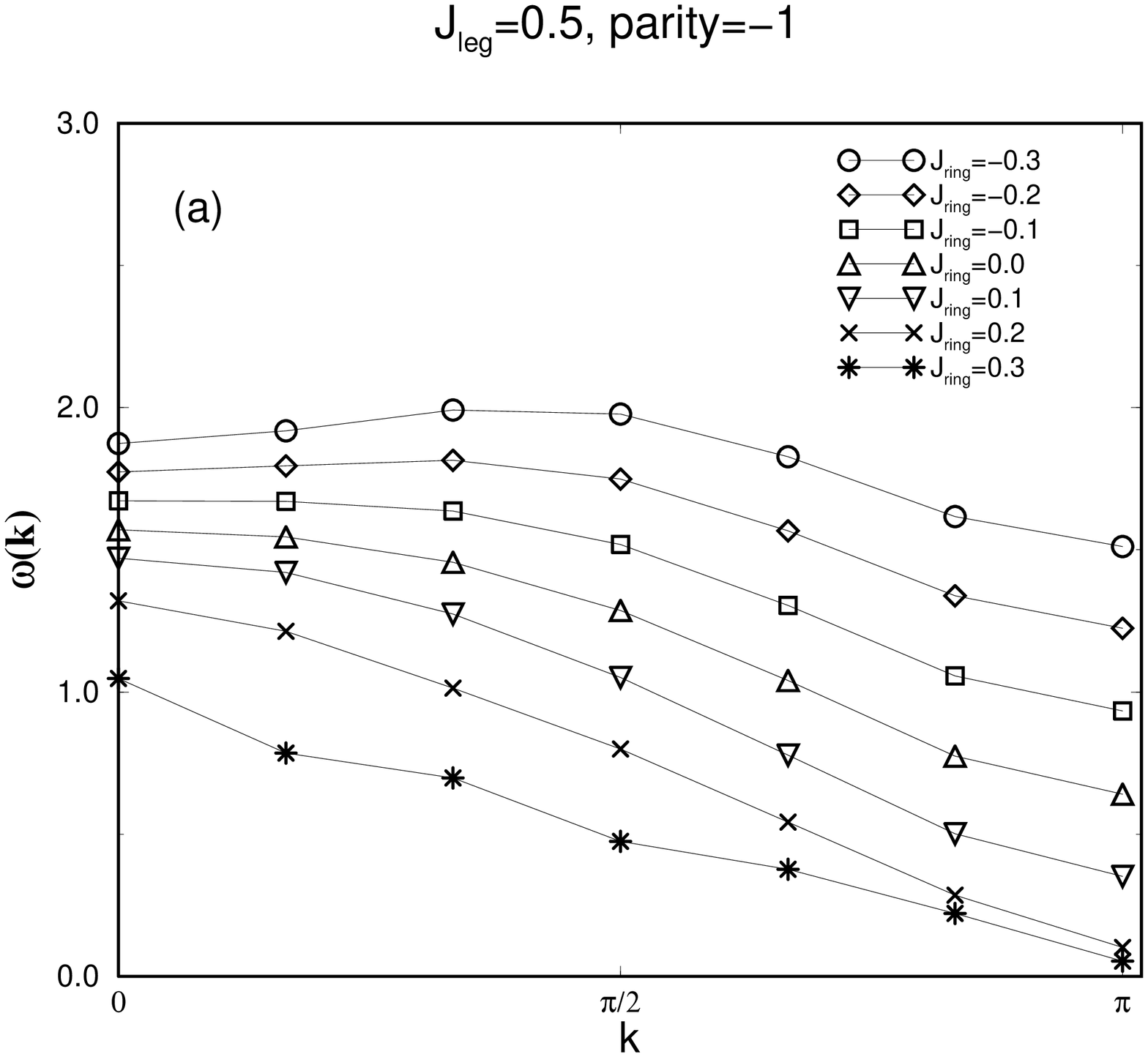,scale=0.3,angle=-90}
\centering\epsfig{file=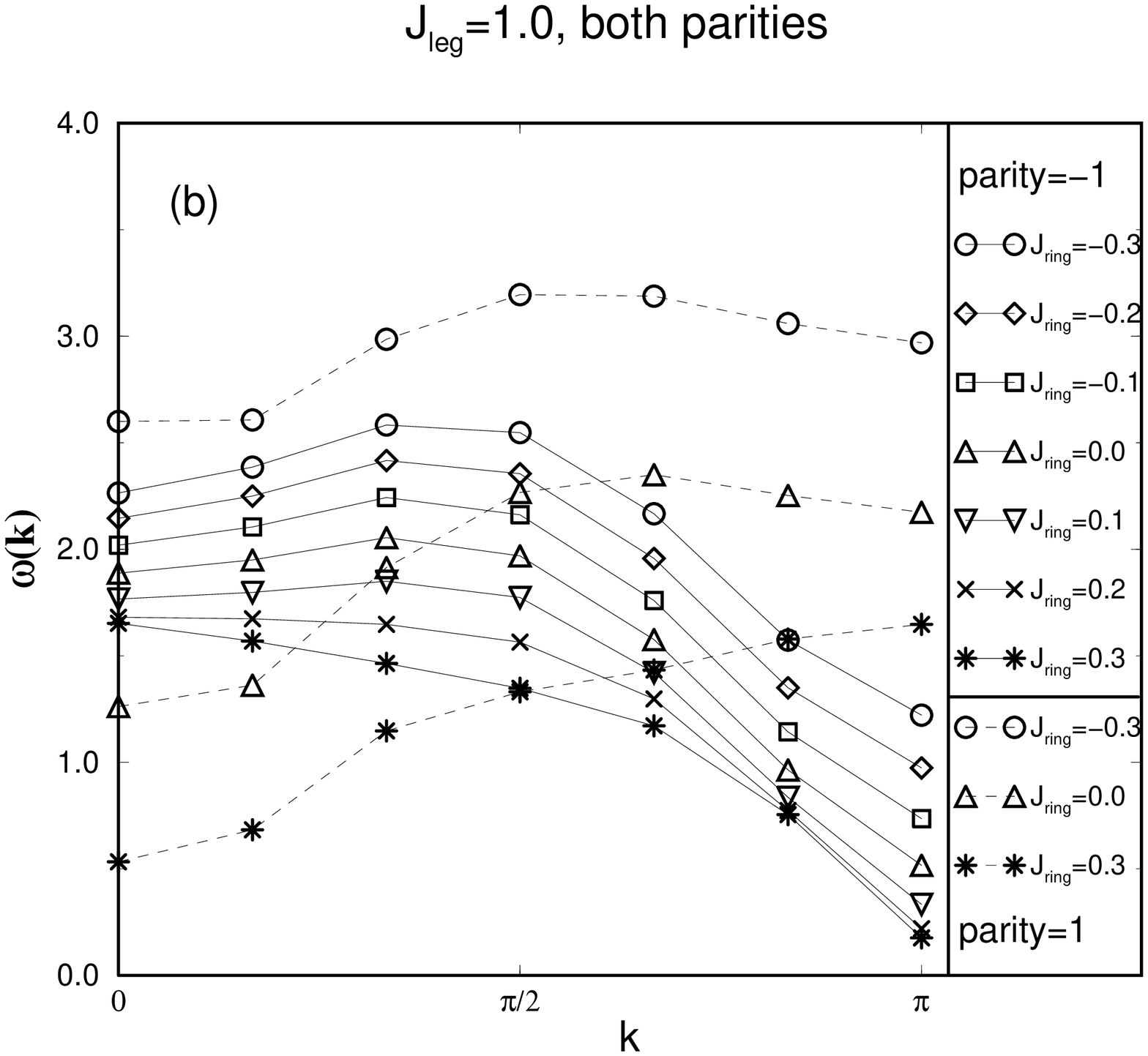,scale=0.3,angle=-90}
\centering\epsfig{file=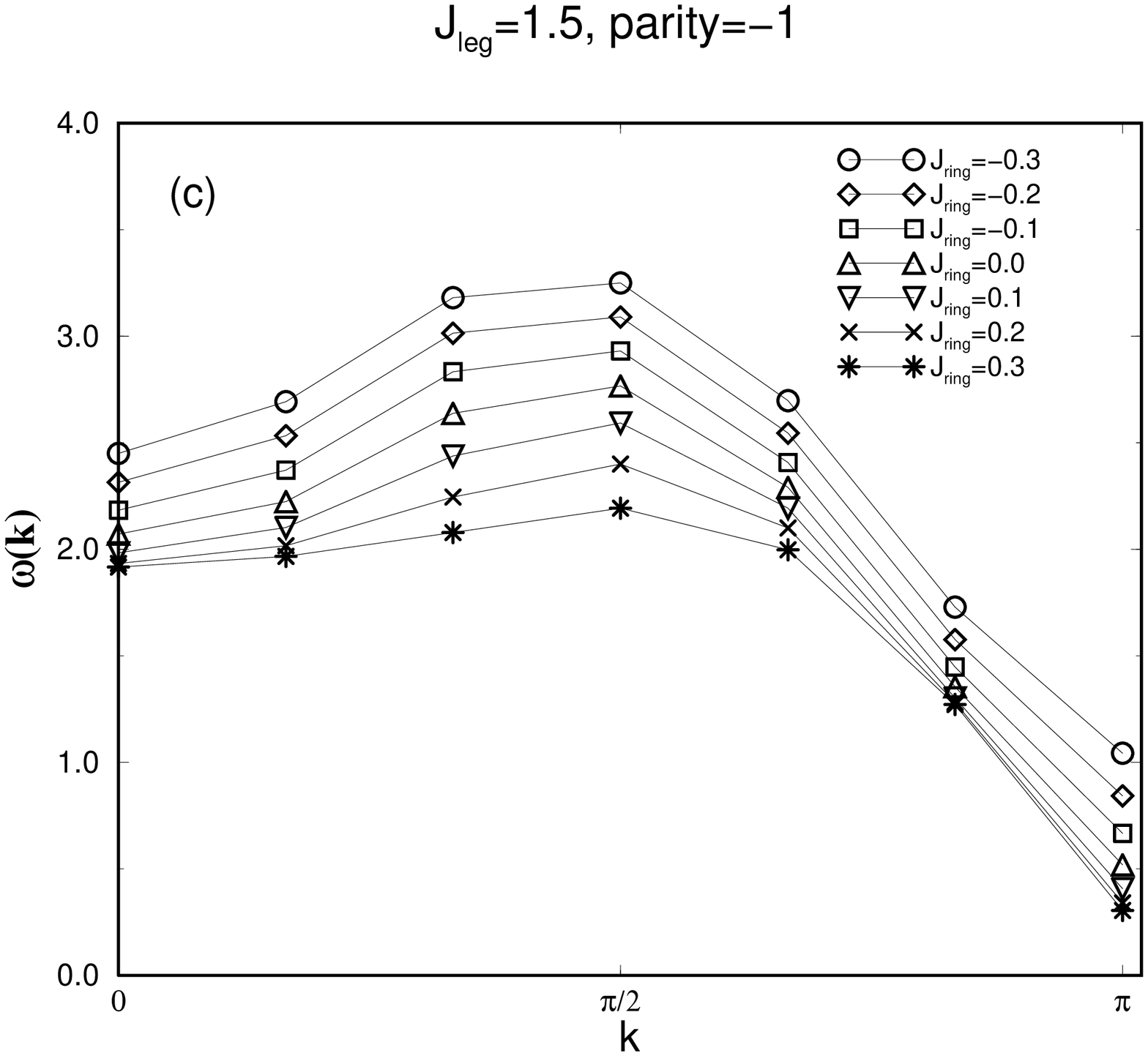,scale=0.3,angle=-90}
\centering\epsfig{file=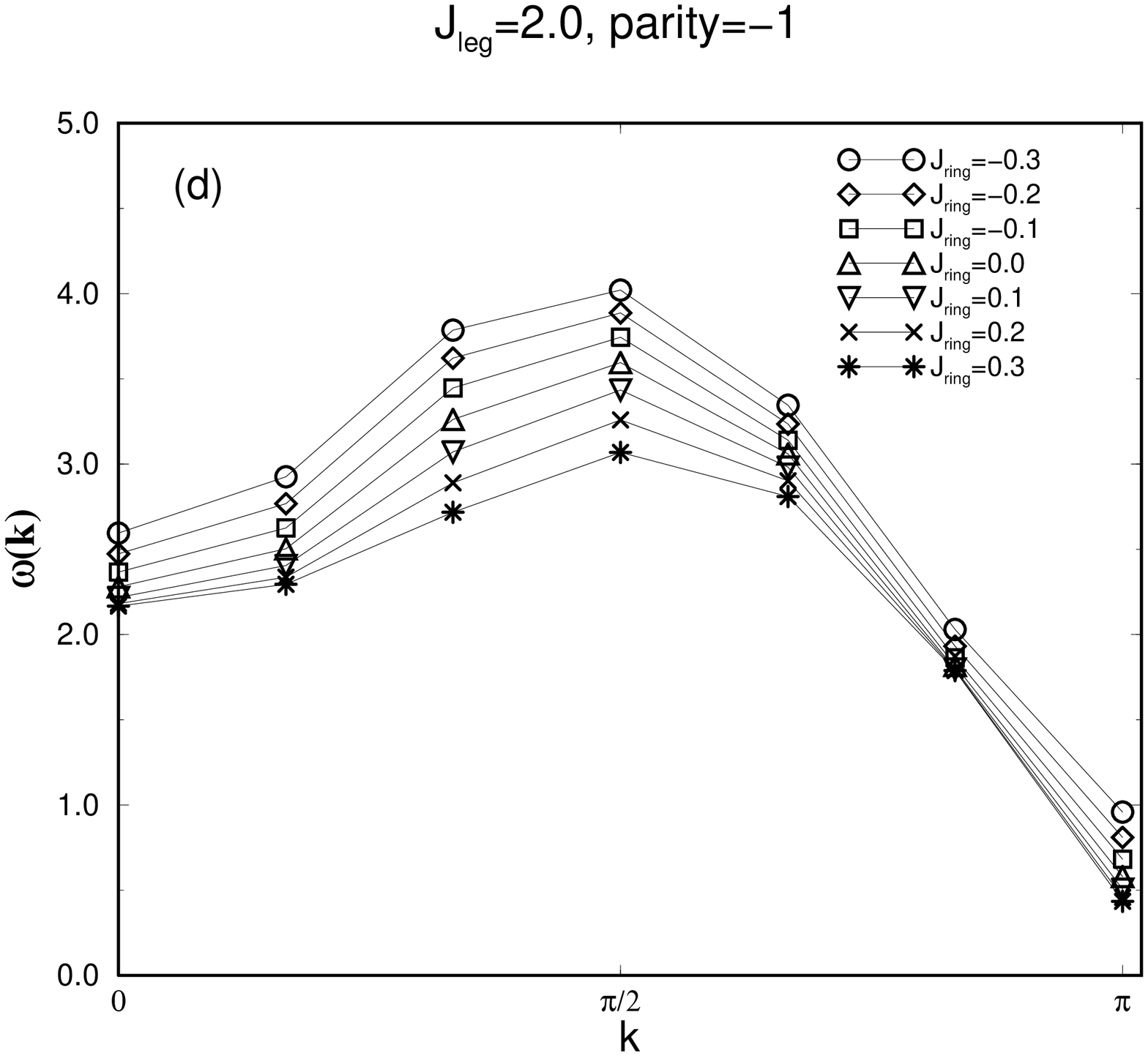,scale=0.3,angle=-90}
\vspace{0.3cm}
\caption{Ladder spectra (lowest negative parity triplet) for $J_{ring} =
-0.3 \dots +0.3$ and $J_{leg}$ = 0.5 (a), 1.0 (b), 1.5 (c) and 
2.0 (d). For $J_{leg}$ = 1.0 the lowest positive parity excitation (singlet)
is also given.}                 
\label{fig:spectra1}
\end{figure}

\begin{figure}
\centering\epsfig{file=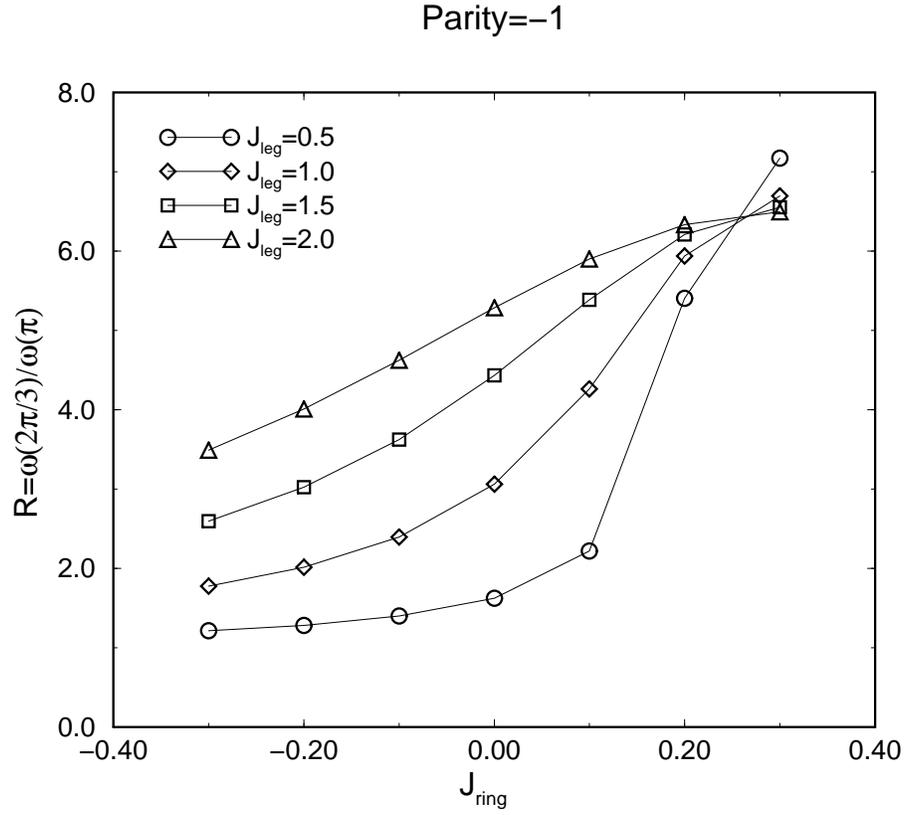,scale=0.57,angle=-90}
\vspace{0.3cm}
\caption{Ratio R (see eq. (5)) characterizing the steepness of the dispersion 
near wavevector $k=\pi$.} 
\label{fig:ratio}
\end{figure}

\begin{figure}
\centering\epsfig{file=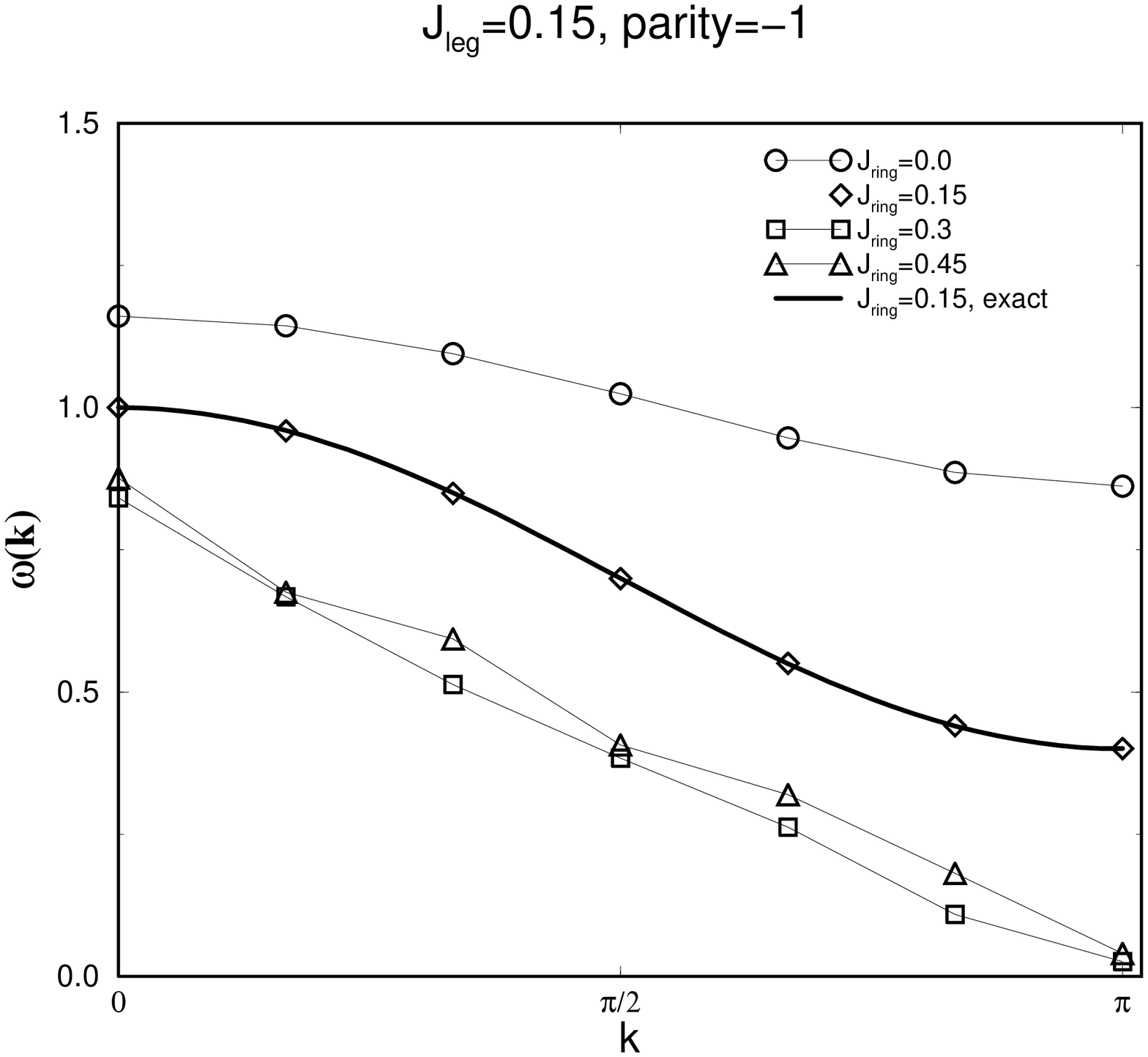,scale=0.57,angle=-90}
\vspace{0.3cm}
\caption{Basic negative parity triplet for $J_{leg}^{bl}=0.15$ and $J_{ring}=0
\dots 0.45$. For $J_{ring}=0.15$ the numerical results reproduce the 
exactly known dispersion curve.}
\label{fig:exact}
\end{figure}

\begin{figure}
\centering\epsfig{file=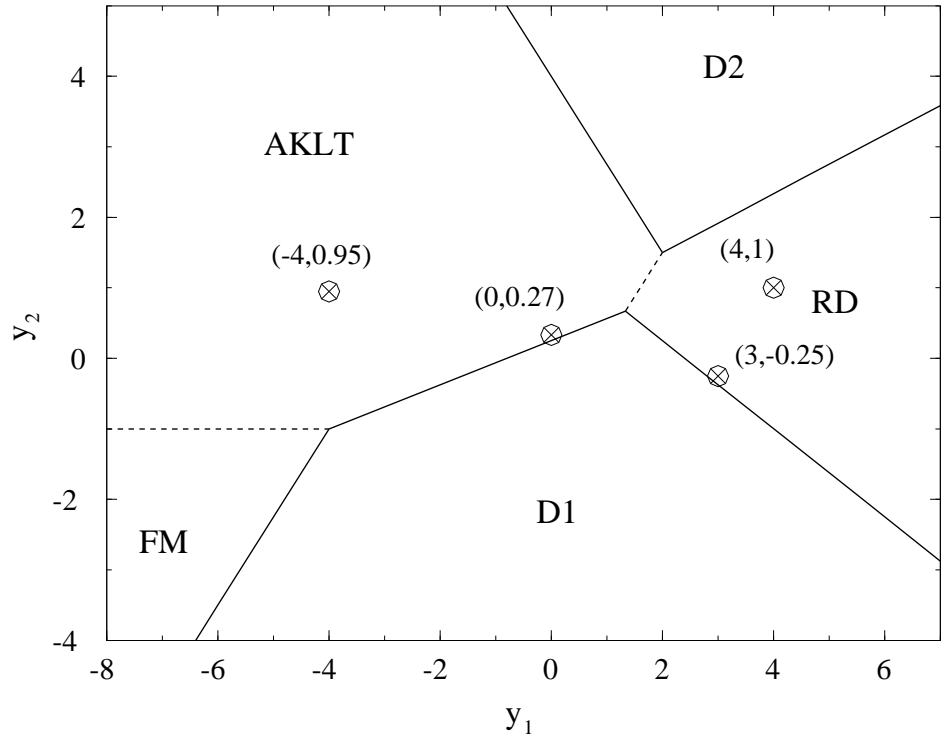,scale=0.57,angle=-90}
\vspace{0.3cm}
\caption{Phase diagram of the generalized Bose-Gayen model 
(see ref. 2)} 
\label{fig:akltphases}
\end{figure}

\begin{figure}
\centering\epsfig{file=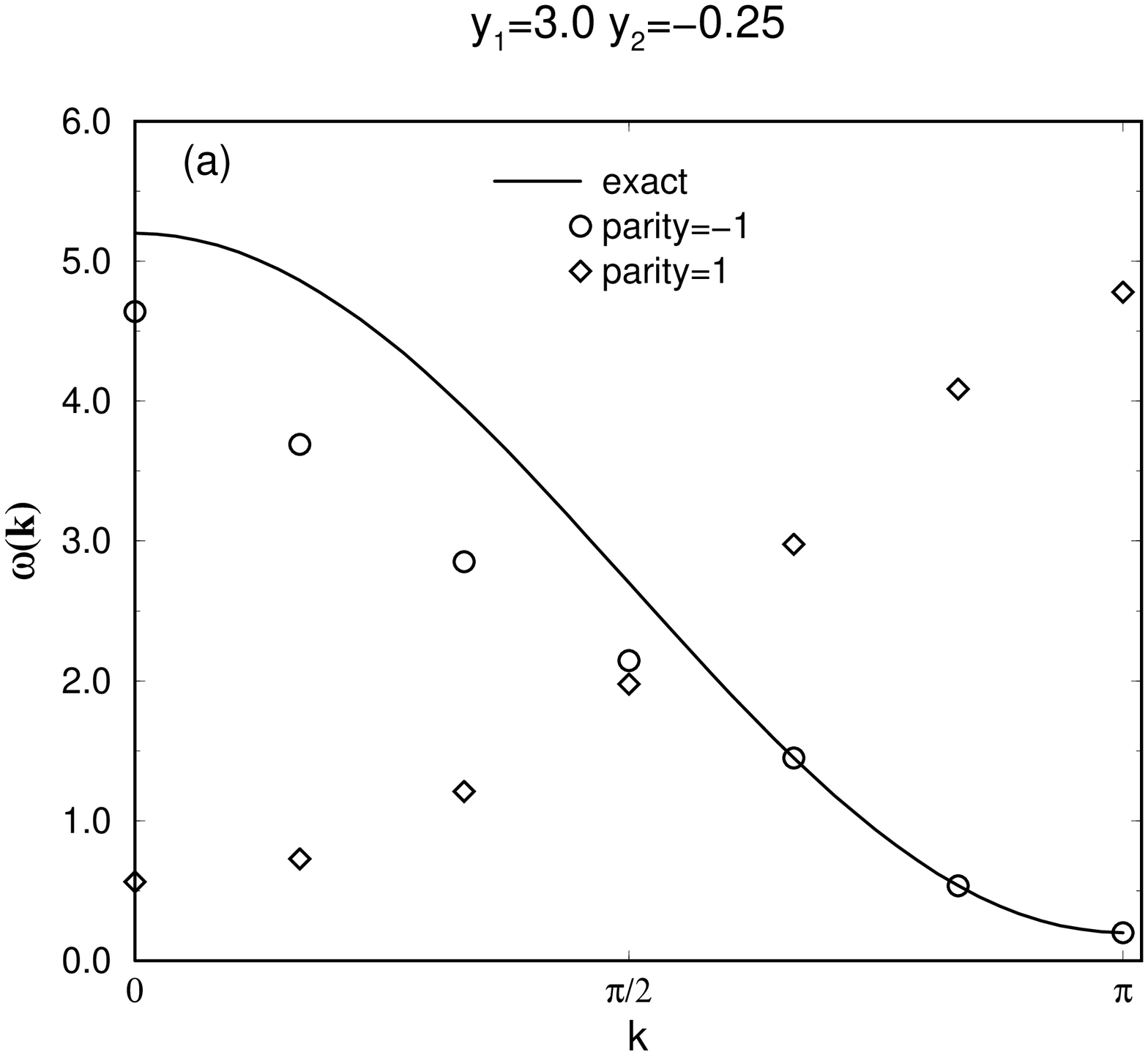,scale=0.3,angle=-90}
\centering\epsfig{file=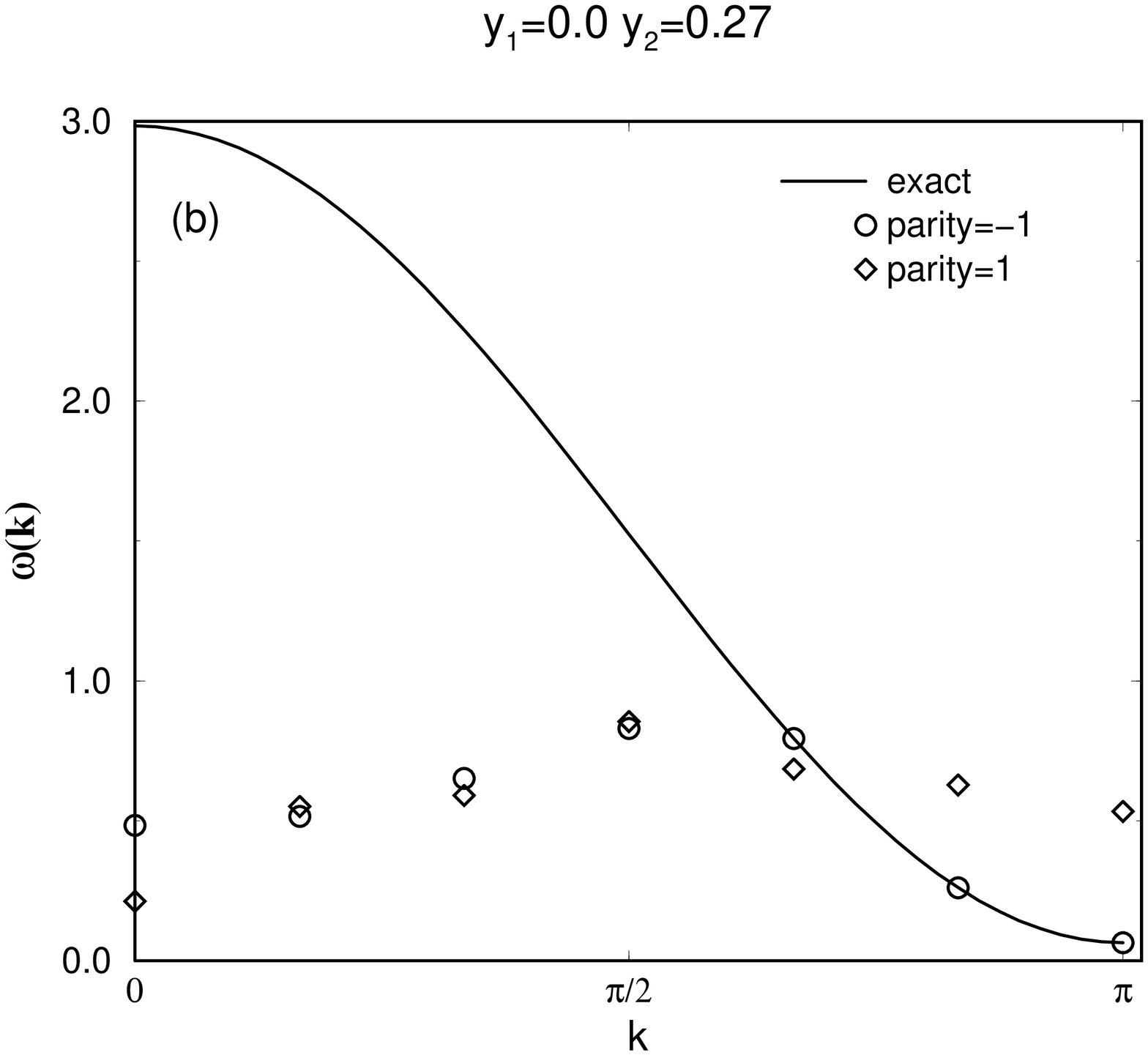,scale=0.3,angle=-90}
\centering\epsfig{file=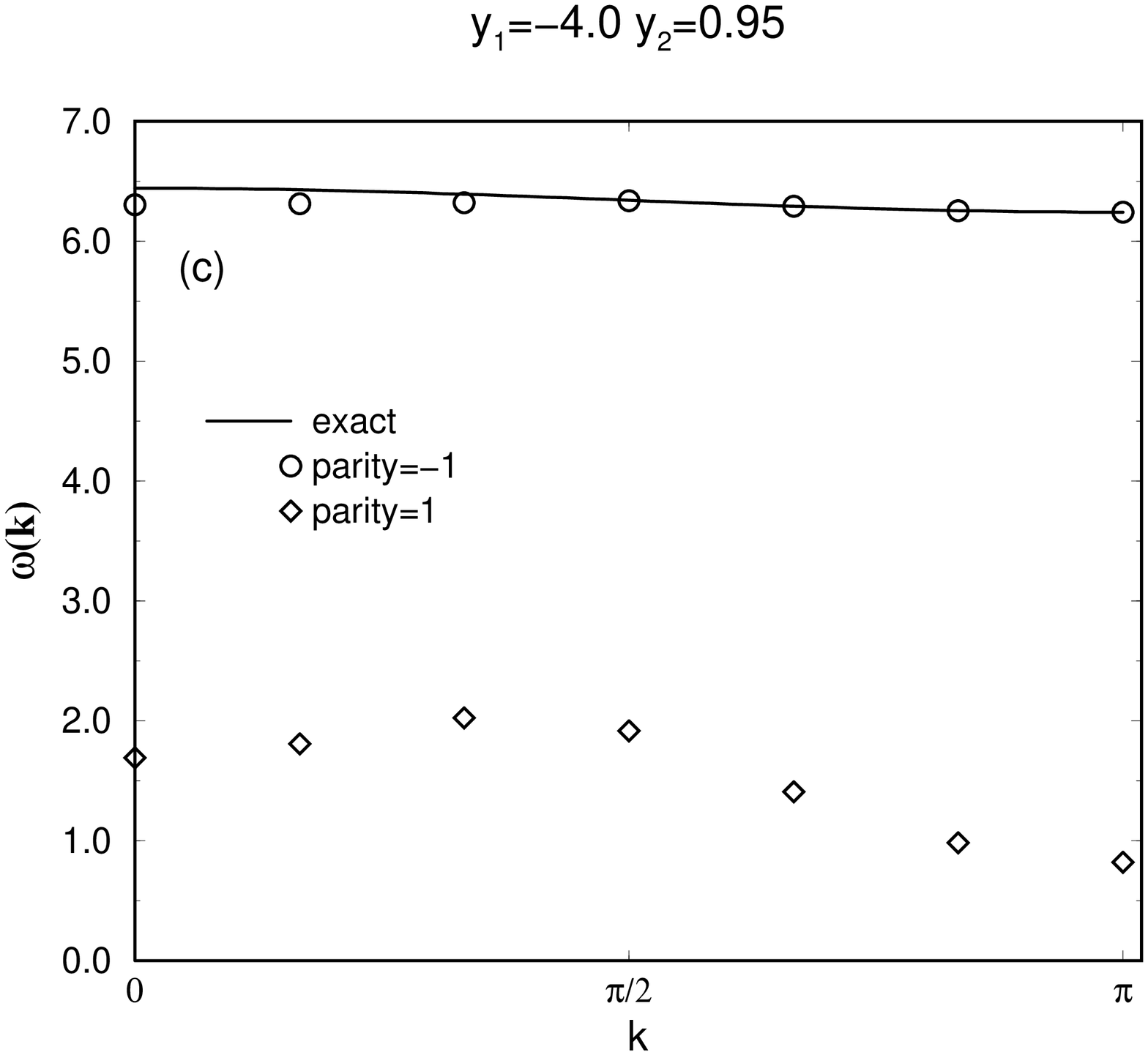,scale=0.3,angle=-90}
\vspace{0.3cm}
\caption{Numerical and exact excitations in the generalized Bose-Gayen model:
The exactly known triplet dispersions are numerically reproduced as lowest 
negative parity excitations for $k > k_{min}$: (a): $k_{min} \ge 2\pi/3$,
(b): $k_{min} \ge 2\pi/3$, (c): $k_{min} \ge \pi/2$.} 
\label{fig:spectra2}
\end{figure}

\end{document}